\documentclass[osajnl,twocolumn]{revtex4}
\usepackage{graphicx}
\graphicspath{{pict/}{}}

\usepackage{bm}% bold math

\newcounter{Fig}

\newcommand{\be}{\begin{equation}}
\newcommand{\ee}{\end{equation}}

\begin{document}

\title{Discrete and surface solitons in photonic graphene nanoribbons}

\author{Mario I. Molina$^{1,2}$ and Yuri S. Kivshar$^3$}

\affiliation{$^1$Departmento de F\'{\i}sica, Facultad de Ciencias,
Universidad de Chile, Casilla 653, Santiago, Chile\\
$^2$Center for Optics and Photonics, Universidad de Concepci\'{o}n, Casilla 4016, Concepci\'{o}n, Chile\\
$^3$Nonlinear Physics Center, Research School of Physics
and Engineering, Australian National University, Canberra ACT 0200,
Australia}

%\date{\today}

\begin{abstract}
We analyze localization of light in honeycomb photonic lattices restricted in one dimension
which can be regarded as an optical analog of (``armchair'' and ``zigzag'') graphene nanoribbons.
We find the conditions for the existence of spatially localized states and discuss
the effect of lattice topology on the properties of discrete solitons excited inside the lattice
and at its edges. In particular, we discover a novel type of soliton bistability, the so-called geometry-induced bistability, in the lattices of a finite extent.
\end{abstract}

%\ocis{190.4420; 190.5530; 190.5940}

\maketitle

The studies of a monolayer of graphite sheet, called {\em graphene}, have attracted
growing attention due to many interesting transport properties of electrons~\cite{review}. Moreover,
semi-infinite graphene and finite stripes of graphene (called {\em graphene nanoribbons}) with zigzag
edges support peculiar electronic states with nearly flat dispersion.

The interesting phenomena in graphene structures are not limited to the electronic systems, and they
have direct analogs in the physics of photonic crystals~\cite{benisty,japan,prl} and photonic
lattices~\cite{moti_1,moti_2}. As a matter of fact, many of the phenomena are generic to honeycomb lattices and can apply to electromagnetic waves in photonic lattices, quasi-particles in graphene, and cold atoms in optical lattices.

All the problems considered for electronic properties of graphene are linear, and no nonlinear effects were discussed so far. However, the photonic analogy suggests not only the study of nonlinear effects in graphene-like structures such as spatially localized nonlinear modes~\cite{discrete_1,discrete_2,discrete_3}, but also a possibility of direct experimental verifications of many of the predicted phenomena, for both hexagon and honeycomb two-dimensional lattices~\cite{moti_1,our_prl}.

In this Letter we employ the analogy with graphene nanoribbons, and study localization of light in honeycomb photonic lattices of a finite extent, an optical analog of graphene nanoribbons. We find the conditions for the existence of spatially localized states and reveal the substantial influence of the lattice topology (i.e. ``armchair'' or ``zigzag'') on the properties of discrete solitons excited
inside the lattice or at its edges.

\begin{figure}[h]
\noindent
\includegraphics[scale=0.65]{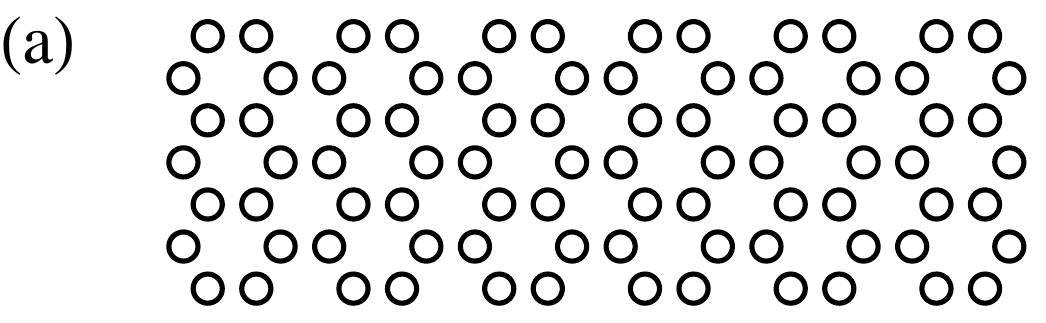}\\
\includegraphics[scale=0.65]{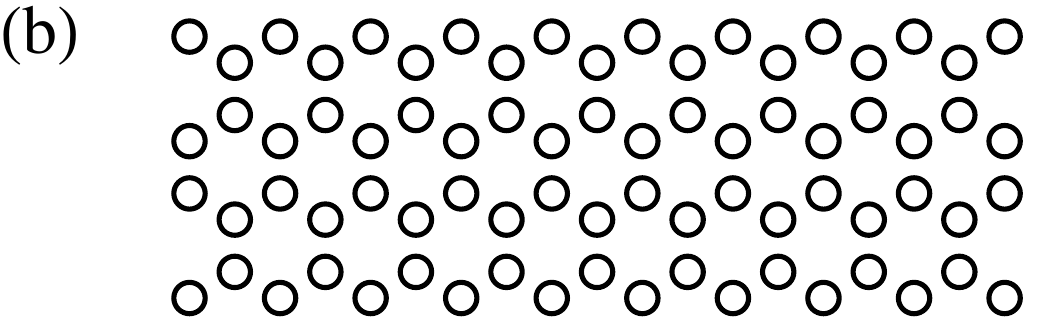}
\caption{Schematic of optical lattice nanoribbons with (a) armchair and (b) zigzag graphene geometries.}
\label{fig1}
\end{figure}

We consider a two-dimensional honeycomb photonic lattice with a finite extension in one dimension,
an optical analog of the graphene nanoribbons. Such photonic stripes can have two distinct geometries,
which can be classified by employing the graphene terminology as ``armchair'' and ``zigzag'' structures,
as shown in Figs.~1 (a,b), respectively.  In the framework of the coupled-mode theory, the electric field
${\cal E}({\bf r})$ propagating along the waveguides can be presented as
a superposition of the waveguide modes, ${\cal E}({\bf
r})=\sum_{\bf n} {\cal E}_{\bf n} \phi({\bf r}-{\bf n})$, where
${\cal E}_{{\bf n}}$ is the amplitude of the (single) guide mode
$\phi({\bf r})$ centered on site with the
lattice number ${\bf n}= (n_1,n_2)$.  The evolution equations
for the modal amplitudes $E_{\bf n}$ take the well-known form,
\be i {d {\cal E}_{\bf n}\over{dz}} + V \sum_{n_1,n_2}{\cal
E}_{\bf m} + \gamma |{\cal E}_{\bf n}|^2 {\cal E}_{\bf n} =
0,\label{eq:1}
\ee
where ${\bf n}$ denotes the position of a guide center, and $V$ is the coupling
in the lattice. The nonlinear parameter $\gamma$ is normalized to $1$
for the focussing nonlinearity.

Next, we analyze the stationary localized modes of Eq.~(\ref{eq:1})
in the form ${\cal E}_{\bf n}(z) = E_{\bf n} \exp(i \beta z)$, where the
amplitudes $E_{\bf n}$ satisfy the nonlinear difference equations,
\be %
 -\beta E_{\bf n} + V \sum_{n_1,n_2}E_{\bf m} + \gamma |E_{\bf
n}|^2 E_{\bf n} = 0
\label{eq:2}
\ee

\begin{figure}[h]
\noindent
\includegraphics[scale=0.2]{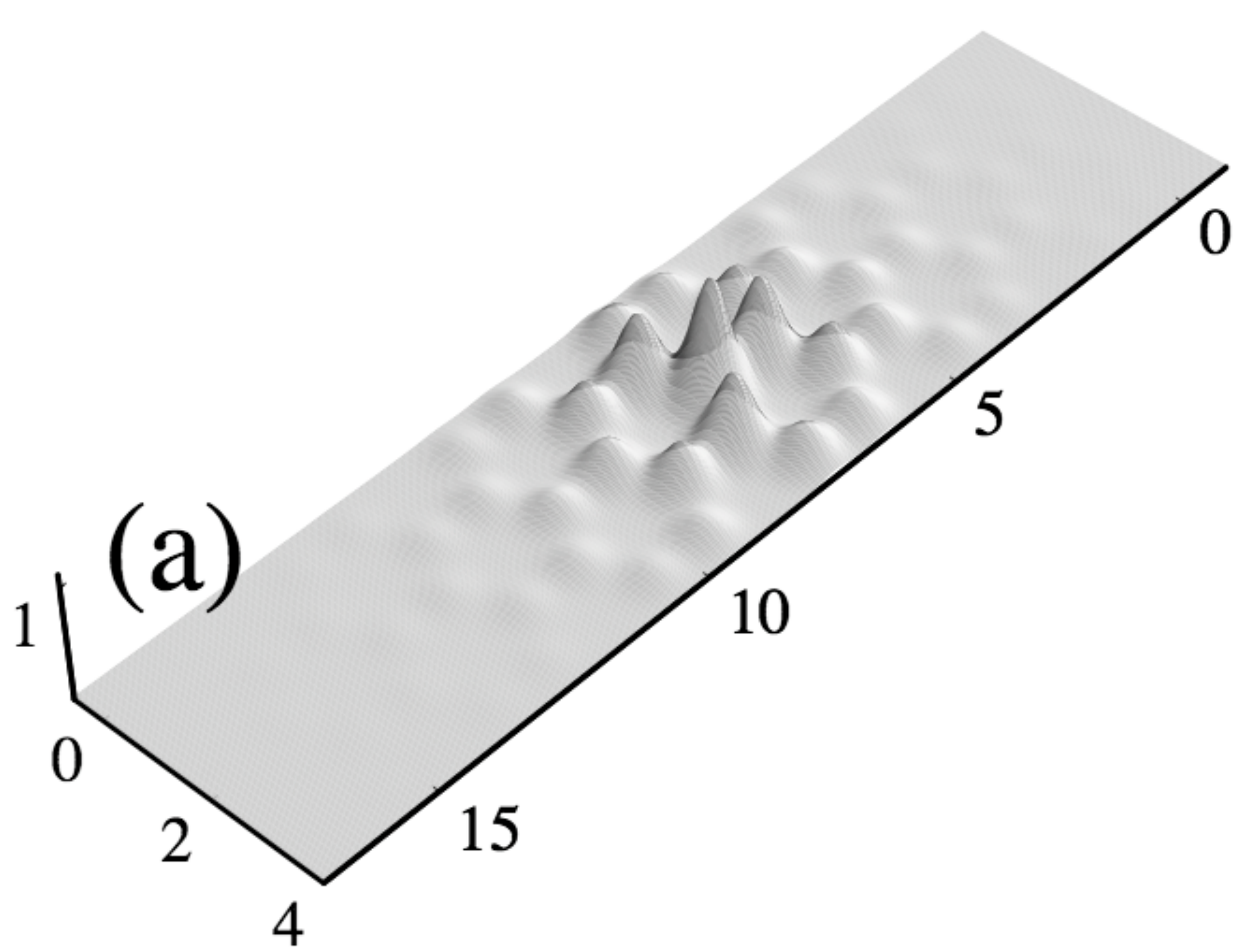}
\includegraphics[scale=0.25]{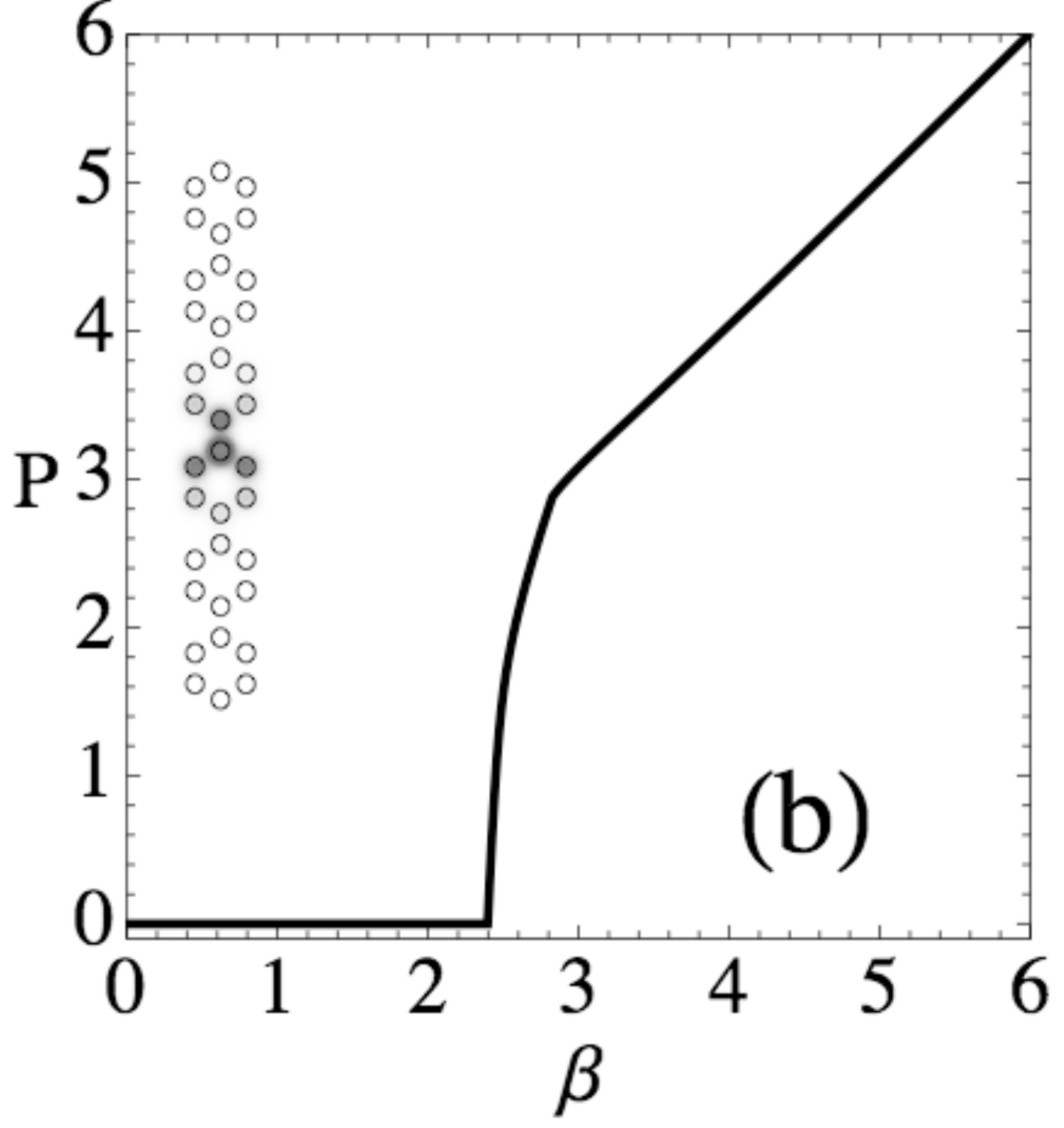}\\
\includegraphics[scale=0.25]{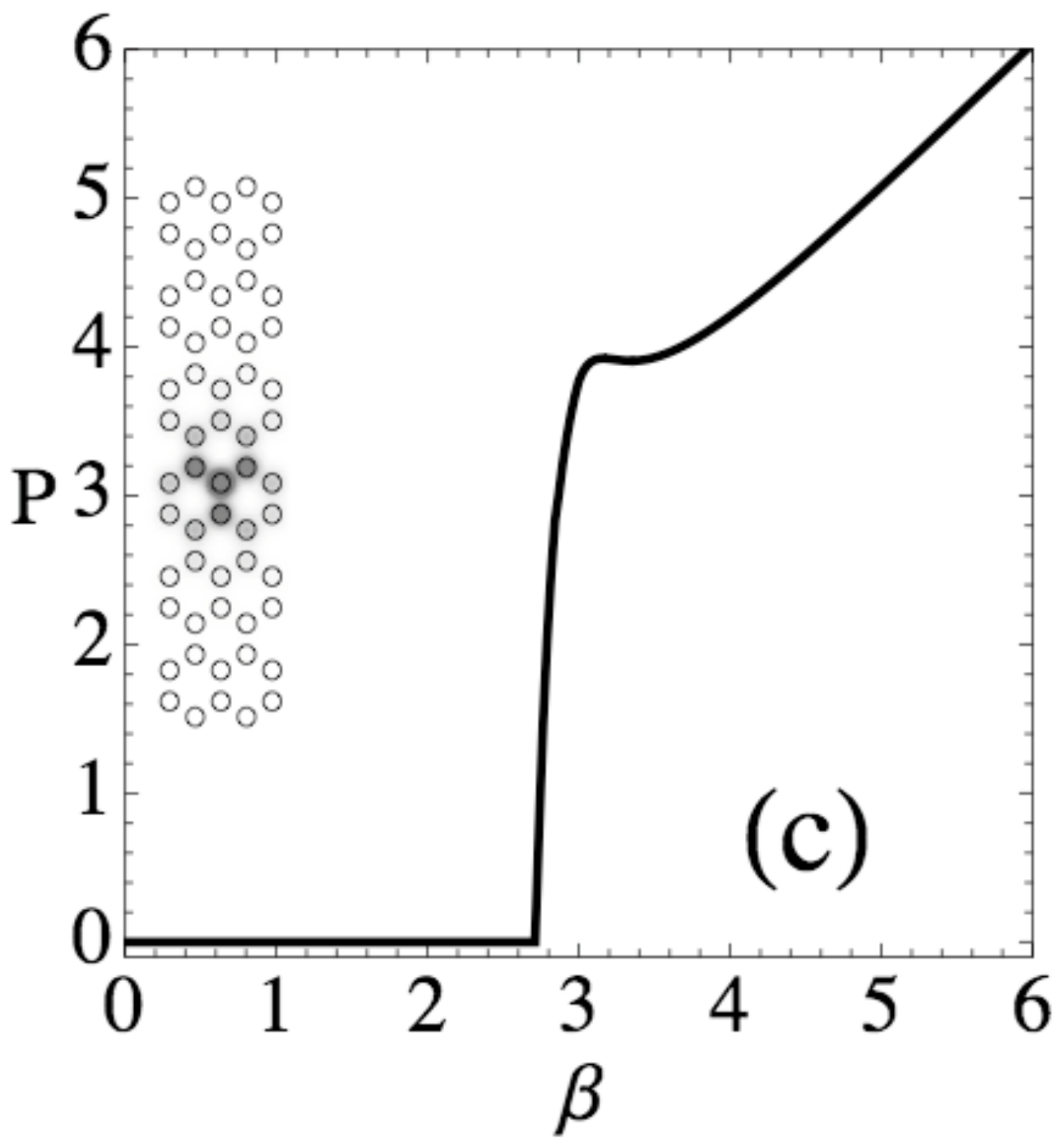}\hspace{0.35cm}
\includegraphics[scale=0.25]{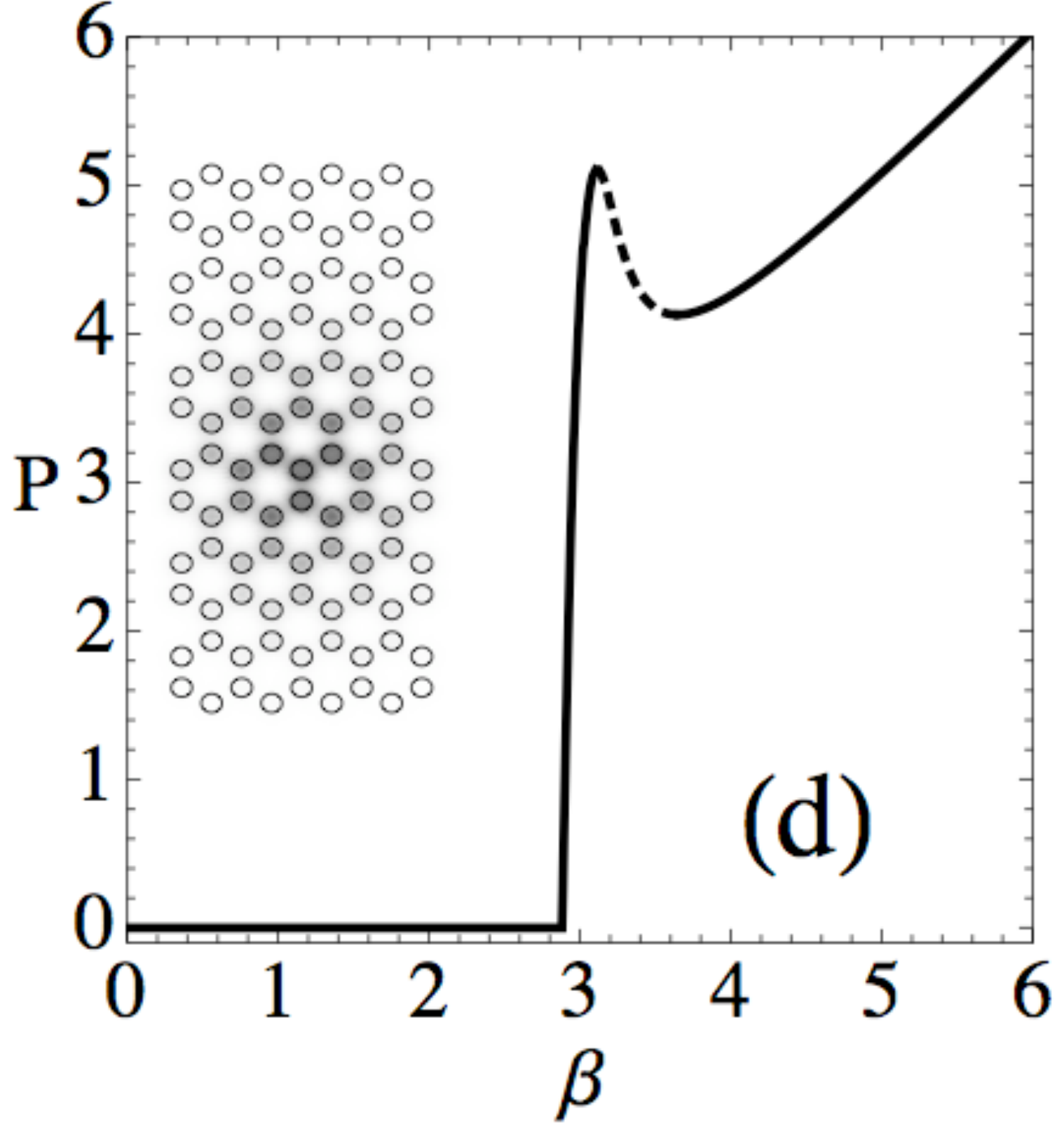}
\caption{Optical localized modes for the graphene nanoribbon with the armchair geometry. (a) Three-dimensional intensity profile of a typical localized mode in a width $3$ nanoribbon (see Fig.2(c)). (b-d) Power vs. propagation constant for the fundamental localized modes in a nanoribbon with the cross-width of (b) one, (c) two, and (d) four elementary cells. The insets show the corresponding structures with the shading of the intensity distribution corresponding to the nonlinear mode at $\beta\sim 3$.}
\label{fig2}
\end{figure}

We consider a nonlinear case for which the linear regime can be achieved in the limit $P \rightarrow 0$ where $P= \sum_{\bf n}|E_{\bf n}|^2$ is the mode power. For a given value of $\beta$,
the system of stationary equations (\ref{eq:2}) is solved numerically by a multidimensional Newton-Raphson scheme. As we are interested in the modes localized inside the structure, we look
for the localized solutions with the maxima near the center decaying quickly along and across the stripe.
In order to visualize the field in the lattice, we present the field as $U(x,y) = \sum_{n,m} C_{n,m} \phi(x-n, y-m)$, where $\phi$ is a guided mode of a single waveguide centered at the site $(n,m)$. For the latter function, we assume a generic form, $\phi(x,y)= \exp [-(x^2 +y^2)/\sigma]$, taking $\sigma =0.1$.
Figure~\ref{fig2}(a) shows an example of the modes localized in the optical graphene stripe.

We find that the results vary substantially depending on the width of the stripe. In both  cases, we find that relatively narrow stripes show the properties of one-dimensional nonlinear chains, where the spatially localized modes do not exist in the linear limit but split off the edge of the continuous spectrum for the power dependence $P(\beta)$. For $P \rightarrow 0$, this curve approaches the value $\beta_{m}$ that coincides with the edge of the linear band. In particular, for the narrow stripe of Fig. 2(b), we find analytically that the linear dispersion is described by two branches, $\beta_{1,2}(k) = V[ 3 \pm 2(1+ \cos k)^{1/2}]^{1/2}$, so that $\beta_m = V(3+2\sqrt{2})^{1/2}$ which for $V=1$ gives $\beta_m \approx 2.414$, corresponding to the cutoff value in Fig.~2(b).

For wider stripes, we observe the appearance of a kink in the power dependence and the corresponding mode bistability [see, e.g., Figs.~\ref{fig2}(d)]. This kink will disappear for much broader (width $8$) stripes, so the lattice of an intermediate extent demonstrates a crossover between one- and two-dimensional lattices. More importantly, the bistable dependence shown by the function $P(\beta)$ demonstrates the first, to the best of our knowledge, example of a geometry-driven bistability of solitons.
Figure 4 shows stable modes on both sides of the bistable curve of Fig. 2(d), obtained by dynamical evolution of the left mode of the branch (Fig. 4(a)), with $\beta=3.0$, perturbed by adding an amount of power exceeding the maximum supported by that branch. After some evolution time, the system
falls into a mode of the right branch (Fig.4(b)), characterized by a sharper
localization, in view of its closer proximity to the anticontinuum limit.

\begin{figure}[h]
\noindent
\includegraphics[scale=0.25]{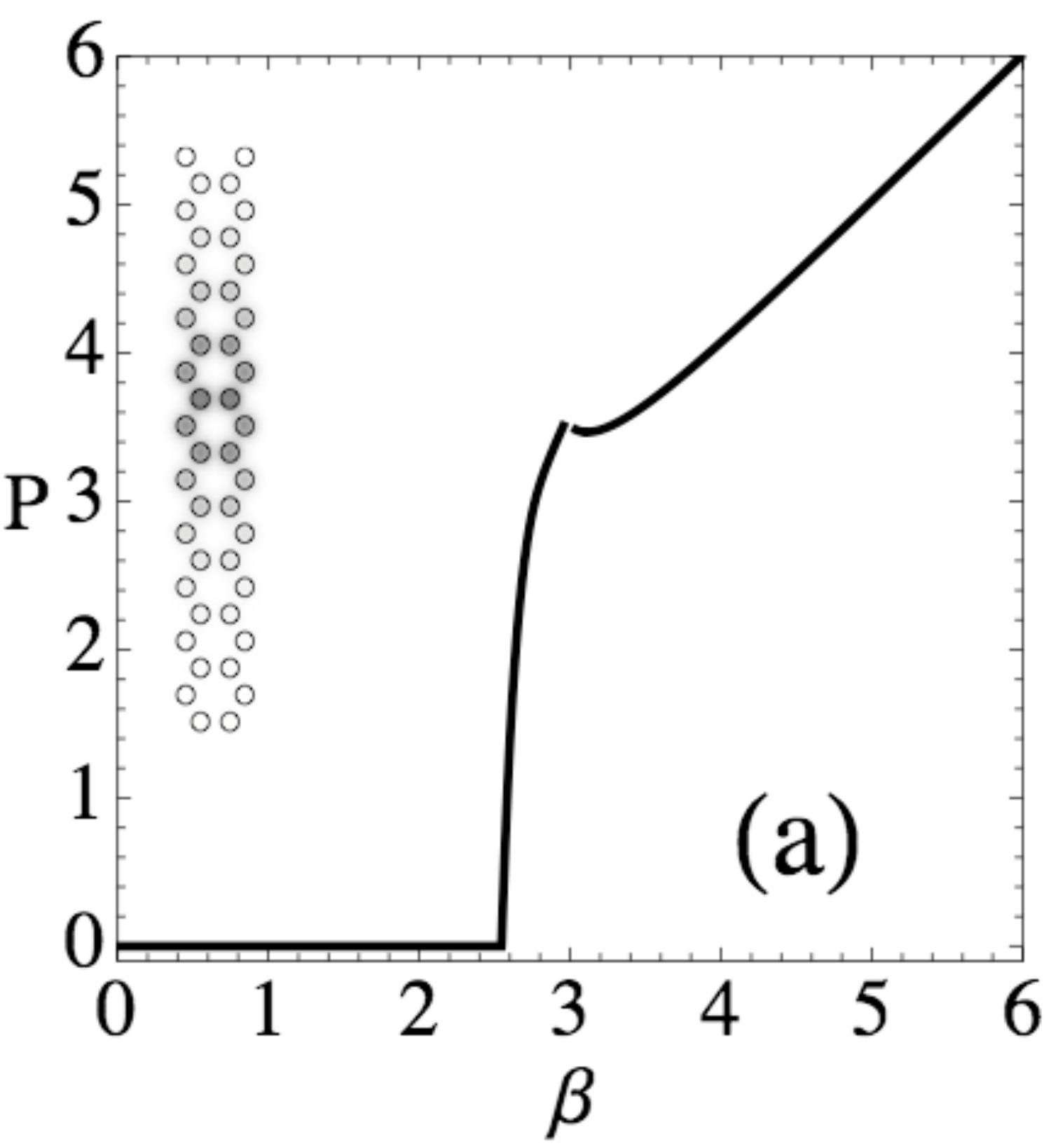}
\includegraphics[scale=0.25]{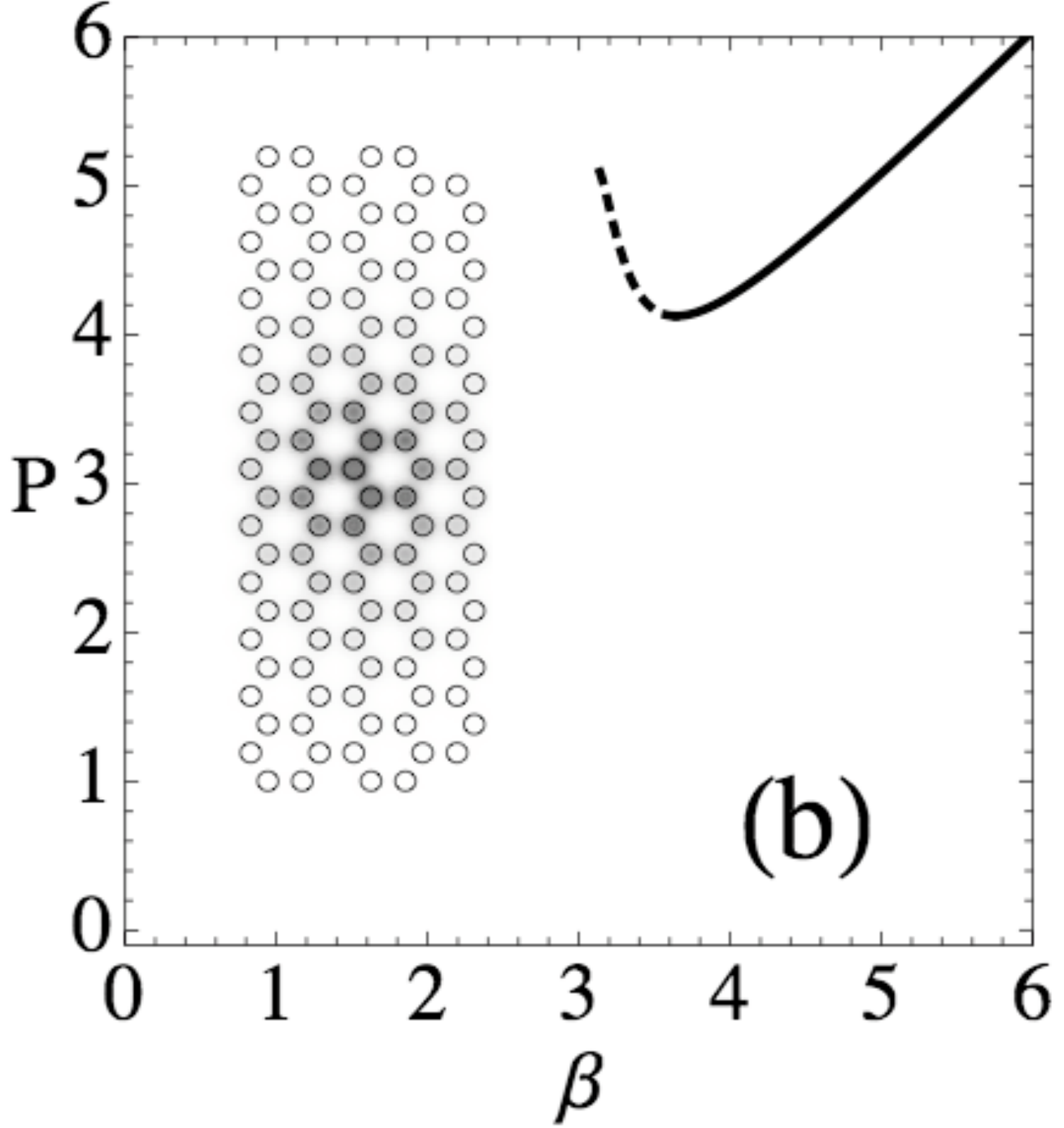}
\caption{Optical localized modes for the graphene nanoribbons with the zigzag geometry. (a)  Power vs. propagation constant for the fundamental localized modes in a nanoribbon with the width of (a) one and
(b) four elementary cells. The insets show the corresponding structures with the shading of the intensity distribution corresponding to the nonlinear mode at $\beta\sim3$.}
\label{fig3}
\end{figure}

Surprisingly, the localized modes in the lattice with the zigzag geometry demonstrate a very different behavior with almost no crossover regime. Figures~\ref{fig3}(a,b) show the power dependencies for two types of ``zigzag'' stripes created of a honeycomb photonic lattice of a finite extent. In the weakly nonlinear regime the localized modes of narrow stripes do show the properties of one-dimensional discrete solitons similar to the modes in the armchair geometry. In particular, for the stripe of Fig.~3(a)
the dispersion can be found in the form $\beta(k) = (V/2) [ 1 + (1 + 16\cos^2k)^{1/2}]$, so that the cutoff value $\beta_m = (V/2)(1 + \sqrt{17})$ which for $V=1$ gives $\beta_m \approx 2.56$. For broader stripes we do not observe the crossover regime, and the power dependence acquire the genuine two-dimensional characteristics, see Fig.~\ref{fig3}(b).

Finally, we analyze surface modes in such photonic structures. Existence of novel types of discrete surface solitons localized in the corners or at the edges of
two-dimensional photonic lattices~\cite{makris_2D,pla_our,pre_2D}
have been recently confirmed by the experimental observation of
two-dimensional surface solitons in optically-induced photonic
lattices~\cite{prl_1} and two-dimensional waveguide arrays
laser-written in fused silica~\cite{prl_2,ol_szameit}. These
two-dimensional nonlinear surface modes demonstrate novel features
in comparison with their counterparts in truncated one-dimensional
waveguide arrays~\cite{OL_george,PRL_george,OL_molina}. In
particular, in a sharp contrast to one-dimensional discrete
surface solitons, the mode threshold is lower at the surface than
in a bulk making the mode excitation easier~\cite{pla_our}.

Here, we employ our photonic nanoribbons and study 
%%%%%%%%%%%%%%%%%% 
\begin{figure}[t]
\noindent
\includegraphics[scale=0.4]{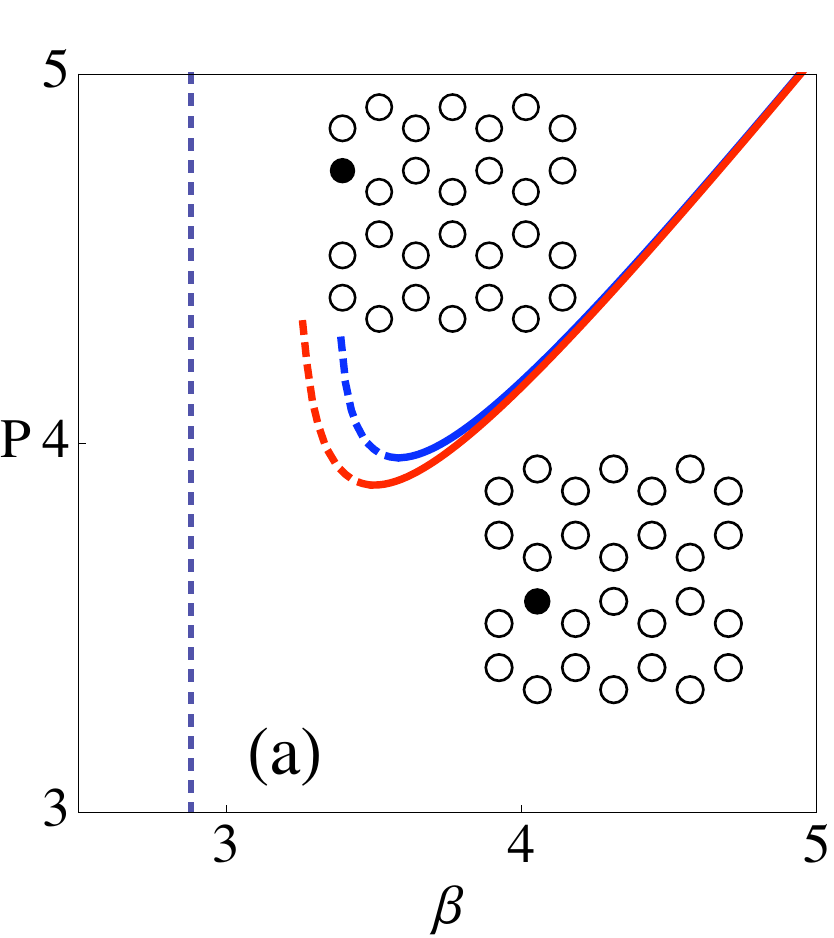}
\includegraphics[scale=0.4]{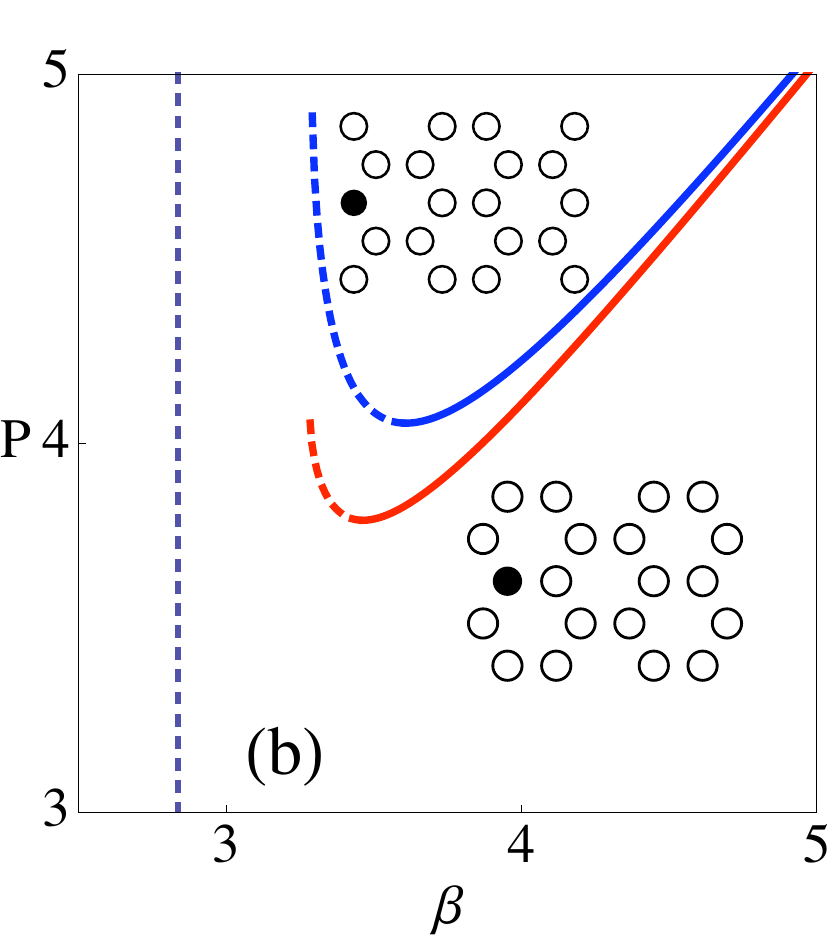}
\caption{Dynamical transition between the two stable localized modes on both sides of the bistable branches of Fig.2(d). After vesting enough power on initial mode with small propagation constant (a), the system evolves dynamically to mode with larger propagation constant (b).}
\label{fig4}
\end{figure}
%%%%%%%%%%%%%%%%%
%%%%%%%%%%%%%%%%%%
\begin{figure}[h]
\noindent
\includegraphics[scale=0.225]{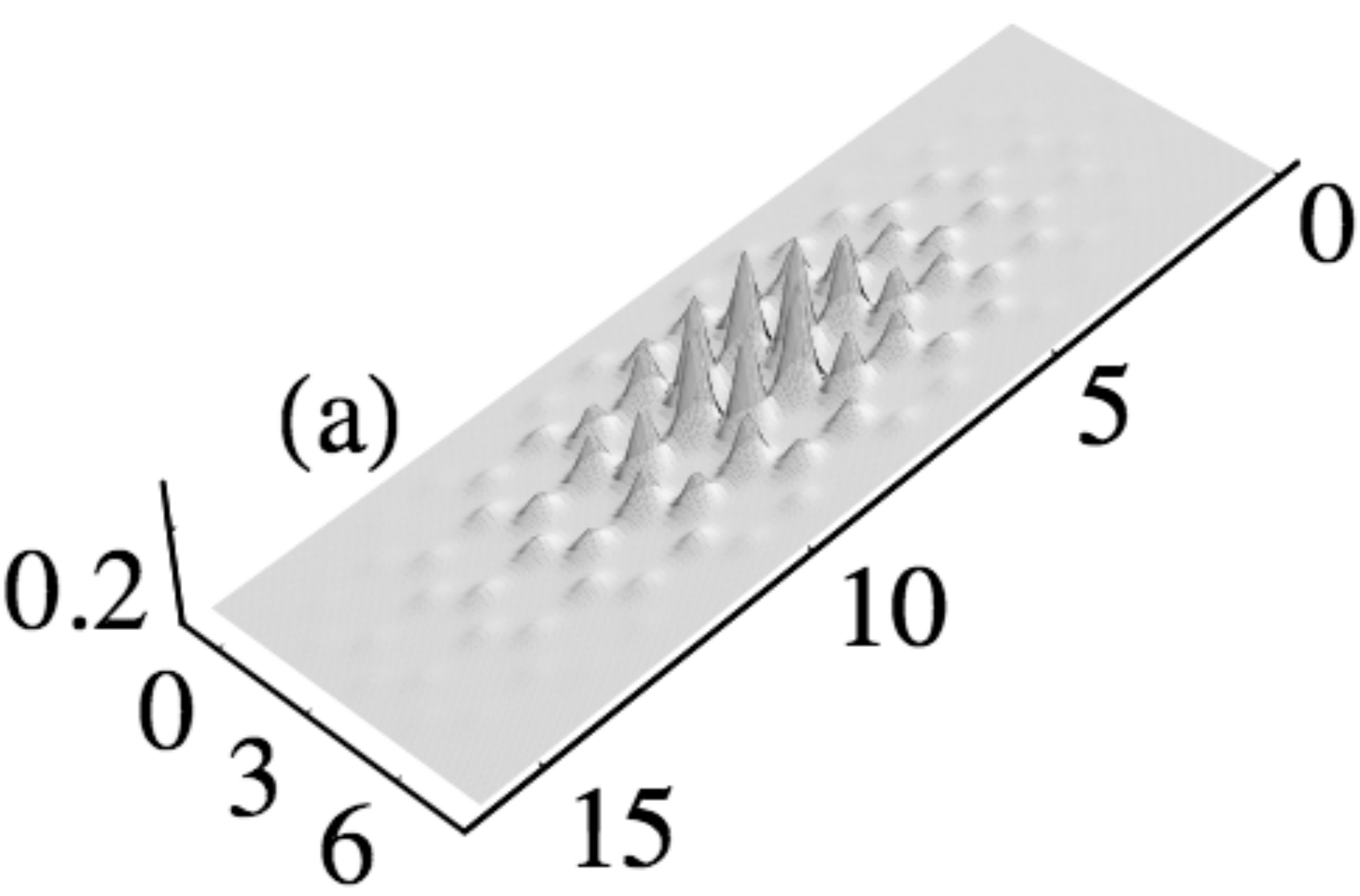}\hspace{0.0cm}
\includegraphics[scale=0.225]{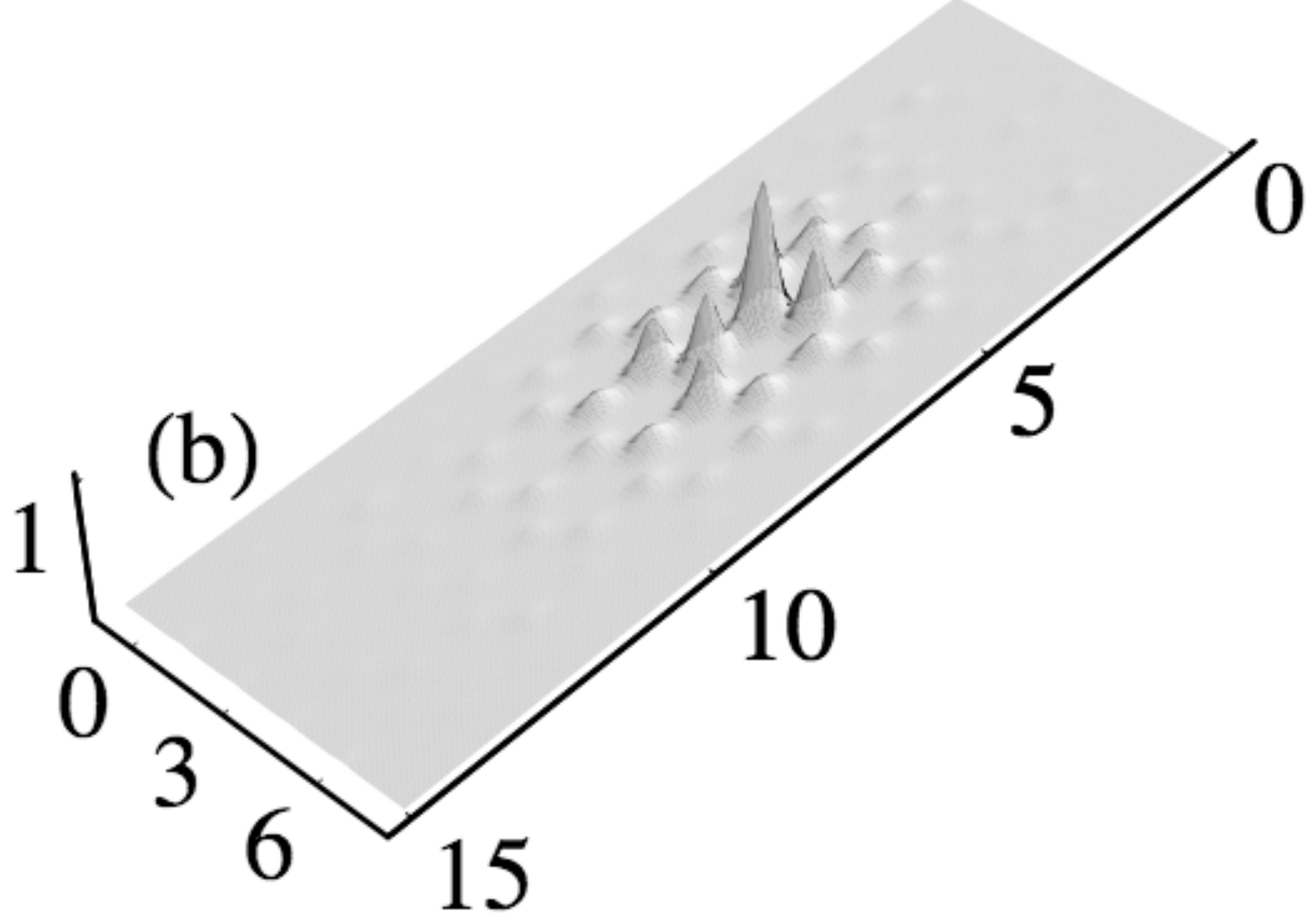}
\caption{Examples of surface modes in a honeycomb optical lattices. (a,b) Power vs propagation constant for for a nanoribbon of width $3$ in the armchair and zigzag geometries, respectively. The upper (lower) curve refers to a mode centered at a boundary site with two (three) nearest neighbors, as shown in the insets. The solid (dashed) curves denotes stable (unstable) portions while the vertical dashed line denotes the position of the linear band.}
\label{fig5}
\end{figure}
%%%%%%%%%%%%%%%%
localization of light at the edge of the lattice. We reveal that the effectively one-dimensional nature of the waveguide created in a two-dimensional lattice
leads to the localized surface modes which resemble the properties of the discrete surface
solitons in a waveguide arrays~\cite{OL_george,PRL_george,OL_molina}.

Figures~\ref{fig5}(a,b) show several examples of low-order nonlinear surface
modes, for both armchair and zigzag geometries of the honeycomb photonic lattice, respectively. These modes do not have their linear counterpart and require a threshold power for their excitation. The stability analysis of those surface nonlinear modes show that the well-known Vakhitov-Kolokolov stability criterium seems to hold, so that the branches with the positive slope in Figs.~\ref{fig5}(a,b) describe stable nonlinear surface states.

In conclusion, we have studied localization of light in two-dimensional
finite-size honeycomb photonic lattices, the so-called photonic graphene nanoribbons.
We have revealed an important effect of the lattice geometry on the
existence and properties of spatially localized modes and discrete solitons.
We have demonstrated that the discrete solitons reveal an interesting feature of
the geometry-induced bistability in the lattice of a finite width. Our results
are generic to honeycomb lattices of a different nature, and they can apply not only
to electromagnetic waves in photonic lattices, but also to quasi-particles in graphene
and cold atoms in optical lattices.

This work was supported by Fondecyt (grant 1080374), Programa de Financiamiento Basal de Conicyt (grant FB0824/2008), and by the
Australian Research Council.

\end{document}